# Algebraic Content of Chiral Symmetry and the Strong Transitions of Heavy Mesons

S.R. Beane[†]

Department of Physics, Duke University
Durham, NC 27708-0305

*sbeane@phy.duke.edu*

## ABSTRACT

The algebraic content of chiral symmetry constrains the strong interactions of mesons containing a single heavy quark. We show that the S-wave single-pion transition amplitudes of all heavy meson states are determined as a consequence of the participation of these states in *reducible* multiplets of *unbroken* $SU(2)_L \times SU(2)_R$. We find this representation content by making use of phenomenologically inspired Lie-algebraic sum rules, together with QCD constraints on the heavy meson spectrum in the heavy quark limit. For example, the transition amplitude for the process $P^* \to P\pi$, where $P$ is a $D$ or a $B$ meson, is predicted to vanish. We also consider the Kaon system in light of our general conclusions for $I = \frac{1}{2}$ states.

[†] This work was supported by the
U.S. Department of Energy (Grant DE-FG05-90ER40592).

## 1. Introduction

It is clearly desirable to pinpoint those aspects of the low-energy strong interactions that are consequences of the symmetries of QCD. Chiral perturbation theory ($\chi PT$) provides a systematic way of realizing this objective. However, $\chi PT$ is useful only in processes where all momenta exchanged are much less than $1\,\text{GeV}$. Moreover, the operators in the chiral lagrangian that encode the interactions of the Goldstone bosons with other hadrons enter with undetermined coefficients which must be taken from experiment. One might then wonder: Is there any sense in which the symmetries of the underlying theory constrain these coefficients?

Dispersion sum rules offer a successful method of obtaining relations among coupling constants and masses. The basic idea is as follows. Analyticity relates an analytic function evaluated at a specific point in the complex plane, say $s_0$, to an integral of the function over all values of $s$. If there exists an S-matrix element with soft enough asymptotic behaviour so that no subtractions are required, and which is uniquely determined by symmetry at the point $s_0$, then saturating the dispersion integral with all possible single-particle singularities of the S-matrix element yields an expression of the form $1 = \sum_i \epsilon_i g_i^2$, where $\{g_i\}$ is a set of coupling constants, and $\{\epsilon_i\}$ is a set of numerical coefficients. This sort of relation has proved remarkably successful at constraining coupling constants. Consider the paradigmatic example. In familiar form, the Adler-Weisberger sum rule [1] for the determination of the axial-vector coupling constant, $g_A$, can be expressed as

$$1 = g_A^2 + \frac{2 f_\pi^2}{\pi} \int_0^\infty \frac{d\omega}{\omega} \left[ \sigma_- (\omega) - \sigma_+ (\omega) \right], \qquad (1.1)$$

where $\sigma_\pm$ represents the total cross-section for the scattering of a $\pi^\pm$ on a proton. Resonance saturation of the dispersion integral gives

$$1 = g_A^2 + \sum_\gamma \epsilon_\gamma g_{\gamma N\pi}^2 \qquad (1.2)$$

where $\epsilon_\gamma$ is $-$ or $+$ when the intermediate state, $\gamma$, carries isospin $\frac{3}{2}$ or $\frac{1}{2}$, respectively, and the $g_{\gamma N\pi}$ are suitably normalized coupling constants. Saturation with $\Delta(1232)$ yields $g_A \simeq 1.75$, whereas a more careful analysis [2] which includes the vast number of other intermediate states with an appreciable coupling to the $\pi N$ channel gives $g_A \simeq 1.29$, as compared to the experimental value $g_A^{exp} = 1.25$. This remarkable agreement established the first success of chiral symmetry in explaining low-energy strong interaction phenomena.

On the other hand, from a purely theoretical point of view the pole saturated form of the sum rule, Eq. (1.2), might appear puzzling. Evidently, the left hand side of Eq. (1.2)

is completely determined by symmetry, whereas the many coupling constants which contribute on the right hand side seem to be fine-tuned to special values in order to satisfy the equation. Actually, there is no paradox. The coupling constants are necessarily related by the same symmetry that determines the matrix element at a specific point. Consider instead the effective lagrangian point of view. In the usual non-linear realization of chiral symmetry, the goldstone bosons interact via complicated momentum-dependent operators. Therefore, S-matrix elements obtained from a chiral lagrangian are, in general, polynomials in the energy variables. If one expands the S-matrix in inverse powers of energy, the coefficients of the leading terms must vanish in order that the asymptotic behaviour of the S-matrix be consistent with observation. In general, any arbitrary number of coupling constants can contribute to each coefficient, and so we are again faced with a fine-tuning problem. Since chiral symmetry is the only symmetry in the problem, it should come as no surprise that chiral symmetry relates these coupling constants. These admittedly simplified arguments serve to illustrate that broken symmetries like chirality have *dynamical* consequences —which constrain the S-matrix in a special kinematic region, as well as *algebraic* consequences [3]—which relate coupling constants, and therefore constrain interactions in a kinematic-independent manner. $\chi PT$ is a widely applied method of exploring the dynamical consequences of chiral symmetry. On the other hand, comparatively little work has been done exploring the algebraic consequences of chiral symmetry.

A review of early work on the algebraic consequences of chiral symmetry can be found in Ref. 4. Recently, it was shown that one can reproduce the main results of the non-relativistic quark model —without recourse to the large-$\mathcal{N}_c$ approximation [5]— by means of algebraic sum rules [6]. The chiral representation involving the pion was studied in Ref. 7 with implications for technicolor theories. Algebraic realizations have also been applied to the constituent quark model, for the purpose of explaining why constituent quarks act as bare dirac particles and determining $g_A$ [8], and for investigating the nature of the representation involving the elementary pion that appears in this model [9]. Possible implications for chiral symmetry restoration were pointed out in Ref. 10.

In this paper, we consider how the algebraic consequences of chiral symmetry constrain the strong interactions of mesons containing a single heavy quark. In principle, we would like our conclusions to be pure reflections of the symmetries of QCD. In practice, as is the case in $\chi PT$, we must make use of experimental input. Here experimental information enters in the form of constraints on the asymptotic behaviour of pion scattering amplitudes. The sum rules that emerge from these constraints can be expressed in Lie-algebraic form, and thus have straightforward group theoretical significance. Specifically, the Lie-algebraic sum rules reveal that for each *helicity*, hadrons fall into *reducible* representations of the *unbroken* chiral group [3]. The mixing angles that appear in these reducible representations are also constrained by the Lie-algebraic sum rules. This point of view is attractive since it allows one to view the observed asymptotic behaviour of S-matrix elements as a direct consequence of symmetry, something which is known to be the case in tractable (weakly coupled) gauge theories [11].

The interactions of hadrons that contain a single heavy quark are constrained by the heavy quark symmetries. In the limit in which the heavy quark mass goes to infinity with its velocity held fixed, one can work in an effective theory which, for $N$ heavy quarks, exhibits an $SU(2N)$ spin-flavor symmetry [12,13]. This symmetry has been used to study many properties of heavy hadrons. Here we focus on heavy mesons. In the heavy quark limit, heavy mesons fall into degenerate doublets labelled by the spin and parity of the light quark constituent [14]. Moreover, heavy quark symmetry relates transition amplitudes where heavy mesons emit or absorb pions, and imposes powerful constraints on the heavy meson mass matrix. We will show that these constraints, together with the algebraic content of chiral symmetry, uniquely determine *all* S-wave single-pion transition amplitudes of *all* heavy meson states. This result constitutes a solution of the chiral commutation relations for all helicities. However, as we will see below, this is true only to leading order in heavy hadron $\chi PT$, where there are only S-wave pion transitions. In the spirit of $\chi PT$ and heavy quark symmetry, the usefulness of the algebraic content of chiral symmetry lies in understanding what features of the spectroscopy of heavy mesons can be understood on the basis of *unbroken* chiral symmetry. We find the extent to which the strong interactions of the heavy mesons are determined purely by symmetry to be remarkable. Unfortunately, the strong transitions of the heavy mesons are not well determined experimentally. This, of course, makes it difficult to assess the relevance of our "solution" to the real world. However, we will argue that unbroken chiral symmetry provides a consistent "explanation" for the pattern of what has been observed.

In the Kaon system, on the other hand, there is a great deal of data but no heavy quark symmetry. Nevertheless, we will consider the phenomenology of the low-lying kaons in light of the general conclusions we reach for the helicity zero states of all $I = \frac{1}{2}$ mesons. Kaon phenomenology suggests a picture distinct from what one would expect on the basis of unbroken chiral symmetry in the heavy quark limit. This is no surprise since the spectroscopy of heavy mesons is expected to be quite different from that of Kaons. We will provide a group theoretical interpretation for this difference, based on the representations of unbroken chiral symmetry filled out by these states.

The paper is organized as follows. Section 2 is a review of the technology of the algebraic realizations, and elaborates previous work on the $I = \frac{1}{2}$ sector. In section 3, we review heavy hadron chiral perturbation theory and write down the leading operators responsible for single-pion transitions among the low-lying heavy meson states. We also review some consequences of heavy quark symmetry relevant to the mass matrix. There is little new in these first two sections and yet the material is essential for what follows. In section 4 we consider the algebraic consequences of chiral symmetry in the heavy quark limit. The main results of the paper are in this section. Section 5 is a search for the correct representation content. Specifically, we analyze a chiral quartet composed of the lowest-lying heavy meson states. In section 6 we discuss our results in the context of heavy meson phenomenology, and consider the issue of heavy quark and chiral symmetry breaking. Section 7 is a phenomenological analysis of the K-meson system. In particular, we consider an Adler-Weisberger sum rule for $\pi K$ scattering. Finally, we summarize and conclude, and discuss further applications of this technology which might prove fruitful.

## 2. Mended Chiral Symmetry

First, we focus on QCD with two massless flavors. Consider the process $\pi\alpha \to \pi\beta$, where $\alpha$ and $\beta$ are arbitrary single-hadron states. We assume that this process is determined by the sum of all chiral *tree graphs*. This is certainly true in the large-$\mathcal{N}_c$ limit [5]. However, we will argue below that there are circumstances where chiral tree graphs dominate, independent of large-$\mathcal{N}_c$ arguments. It is straightforward to construct the most general $SU(2) \times SU(2)$ invariant operators which contribute to this process. Of course there are an infinite number of such terms, and only in special cases are all momentum transfers small enough to allow a consistent perturbative expansion. However, it turns out that in a Lorentz frame in which all momenta are *collinear*, the momentum structure of the most general Feynman amplitude involving the sum of all chiral tree graphs becomes transparent[1]. If the scattering amplitude is expanded in inverse powers of energy, the couplings of the heavy particle content to the Goldstone bosons appear at the same order in the energy variable as the term which is protected by chiral symmetry, and which therefore contains no invariant counterterm. The coefficient of this power in the energy variable is easily extracted and takes the form [3]

$$C^{(-)\,\lambda}{}_{\beta j,\alpha i} \propto \{i\epsilon_{ijk}T_k - [X_i^\lambda, X_j^\lambda]\}_{\beta\alpha} \qquad (2.1)$$

for the crossing-odd amplitude and

$$C^{(+)\,\lambda}{}_{\beta j,\alpha i} \propto \{[X_j^\lambda, [X_i^\lambda, m^2]] + [X_i^\lambda, [X_j^\lambda, m^2]]\}_{\beta\alpha} \qquad (2.2)$$

---

1 See Ref. 15 for a concise review.

for the crossing-even amplitude. The roman subscripts are isospin indices, $T_i$ are the isospin matrices, and $\hat{m}^2$ is the mass-squared matrix. The helicity, $\lambda$, is a conserved quantum number in the collinear frame. $X_i$ is an axial-vector coupling matrix, related to the matrix element of the process $\alpha(p,\lambda) \to \beta(p',\lambda') + \pi(q,i)$ in any frame in which the momenta are collinear:

$$M_i(p'\lambda'\beta, p\lambda\alpha) = (4f_\pi)^{-1}(m_\alpha^2 - m_\beta^2)[X_i^\lambda]_{\beta\alpha}\delta^{\lambda'\lambda}. \tag{2.3}$$

So far we have done nothing. In order to extract interesting physics it is necessary to determine experimentally, or otherwise, how the crossing-even and -odd amplitudes behave at high-energy. Here, by "high-energy" we mean energies of order the characteristic scale at which the momentum expansion fails. It is conventional to denote this scale $\Lambda_\chi \sim m_\rho$. At these energies, Regge behaviour —or a mechanism of equivalent efficiency— should kick in in order to soften the bad asymptotic behaviour implied by the derivative expansion. We will require that the sum of all chiral tree graphs behave no worse than what Regge pole theory suggests for the full amplitude [3]. The crossing-odd amplitude is pure $I_t = 1$, for which Regge theory suggests

$$M_{\beta j,\alpha i}^{I_t=1}(\omega,\lambda) \xrightarrow[\omega\to\infty]{} \omega^{\alpha_1(0)-1}, \tag{2.4}$$

where $\alpha_1(0) \simeq 0.5$ is the intercept of the $\rho$ trajectory. If not familiar with this sort of language, one should instead note that there is substantial phenomenological evidence which suggests that the amplitude with $I_t = 1$ satisfies an unsubtracted dispersion relation; witness the successful determination of $g_A$. In fact, we only make use of properties of Regge pole theory which have proved remarkably successful experimentally and so might be said to exhibit QCD behaviour [16]. In any case, $C^{(-)}$ vanishes, leading to the generalized Adler-Weisberger (A-W) sum rule [3]

$$[X_i^\lambda, X_j^\lambda]_{\beta\alpha} = i\epsilon_{ijk}(T_k)_{\beta\alpha}. \tag{2.5}$$

Together with Eq. (2.5), the defining relations

$$[T_i, T_j] = i\epsilon_{ijk}T_k, \tag{2.6}$$

and

$$[T_i, X_j^\lambda]_{\beta\alpha} = i\epsilon_{ijk}(X_k^\lambda)_{\beta\alpha} \tag{2.7}$$

close the chiral algebra and we see that for each helicity, $\lambda$, hadrons fall into representations of $SU(2) \times SU(2)$, *in spite of the fact that the group is spontaneously broken*. However, $X_i$ does not commute with the mass-squared matrix, $\hat{m}^2$, and therefore, in general, these representations are reducible. Consequently, additional constraints are required in order to fix the coefficients which mix the various irreducible representations. It is the reducibility of these representations that "hides" chiral symmetry at low-energies.

Parity conservation has an important consequence. A combined space reflection and $180^o$ rotation about the collinear direction in the process $\alpha \to \beta + \pi$ leads to the selection rule [3]

$$[X_i^{-\lambda}]_{\beta\alpha} = -P_\alpha P_\beta (-)^{J_\alpha - J_\beta}[X_i^\lambda]_{\beta\alpha}, \qquad (2.8)$$

where $P_\alpha$ and $J_\alpha$ are the intrinsic parity and spin of $\alpha$, respectively. The case of zero-helicity is of special interest as there is an additional quantum number to take into account. In the language of symmetries this is so because the $SU(2) \times SU(2)$ algebra has an *endomorphism*, $\Pi$, which leaves the algebra invariant: $\Pi X_i \Pi = -X_i$, and $\Pi T_i \Pi = T_i$ [7]. The eigenvalues of $\Pi$ are *normality*, $\eta_\alpha \equiv P_\alpha(-)^{J_\alpha}$. Eq. (2.8) implies that $[X_i^0]_{\beta\alpha}$ must vanish unless $\alpha$ and $\beta$ satisfy the selection rule $P_\alpha(-)^{J_\alpha} = -P_\beta(-)^{J_\beta}$. So only zero-helicity states of opposite normality communicate by single-pion emission and absorption.

The crossing-even amplitude has both $I_t = 0$ and $I_t = 2$, and so, in general, one must make an assumption[2] about Regge trajectories with exotic ($I = 2$) quantum numbers in order to extract useful information about the mass-squared matrix [3]. Regge pole theory suggests

$$M_{\beta j, \alpha i}^{I_t=2}(\omega, \lambda) \xrightarrow[\omega \to \infty]{} \omega^{\alpha_2(0)}, \qquad (2.9)$$

where $\alpha_2$ is the leading trajectory with exotic quantum numbers. The assumption that such trajectories are absent, $\alpha_2(0) < 0$, allows the derivation of superconvergence relations, which imply that $C_{ji}^{(+)}$ is proportional to $\delta_{ji}$ (i.e. pure $I_t = 0$). It then follows from Eq. (2.5) and a Jacobi identity that [3]

$$[X_j^\lambda, [m^2, X_i^\lambda]]_{\beta\alpha} = -[m_4^2]_{\beta\alpha}\delta_{ij}. \qquad (2.10)$$

This commutation relation implies that the hadronic mass-squared matrix is the sum of a chiral invariant and the fourth component of a chiral 4-vector; i.e. $\hat{m}^2 = \hat{m}_0^2 + \hat{m}_4^2$. When $\hat{m}_4^2$ vanishes, the mass matrix, $\hat{m}^2$, commutes with $X_i$, and hadrons of a given mass form complete chiral multiplets. Since $\hat{m}_4^2$ is related to the $I_t = 0$ amplitude, we see that it is the exchange of Regge trajectories with vacuum quantum numbers that prevents chirality from showing up as an explicit (linear) symmetry of the strong interactions [17]. In modern language, $\hat{m}_4^2$ is the order parameter; when it vanishes, chiral symmetry is restored. This is no surprise since $\hat{m}_4^2$ transforms like the quark condensate, $\langle\bar{\psi}\psi\rangle$[3].

---

2 Note that if the hadrons in consideration carry only isospin $\frac{1}{2}$, with no single-pion transitions to states of higher isospin, any crossing-even amplitude is pure isoscalar in the t-channel, and there is thus no need for a further assumption.

Consider a system of $I = \frac{1}{2}$ mesons. There are clearly no $I = \frac{3}{2}$ mesons. So assuming that there are no pion-meson bound states (presumably an exact statement in the large-$\mathcal{N}_c$ limit), $I = \frac{1}{2}$ mesons have no single-pion transitions to states of higher isospin. Therefore, insofar as unbroken chiral symmetry is concerned, $I = \frac{1}{2}$ mesons behave much like constituent quarks. Hence, our discussion parallels that of Ref. 8. The fundamental difference is that mesons carry integer helicity. The only representations of $SU(2) \times SU(2)$ that contain only a single $I = \frac{1}{2}$ representation of the diagonal isospin subgroup are $(0, \frac{1}{2})$ and $(\frac{1}{2}, 0)$ [8]. So, in general, $I = \frac{1}{2}$ states of definite helicity are linear combinations of any number of these irreducible representations with undetermined coefficients [8]. Mass splitting can only occur as a consequence of mixing between these representations since $\hat{m}^2$ is a sum of $(0,0)$ $(\hat{m}_0^2)$ and $(\frac{1}{2}, \frac{1}{2})$ $(\hat{m}_4^2)$ contributions. In a basis in which all linear combinations of $(0, \frac{1}{2})$ and $(\frac{1}{2}, 0)$ irreducible representations appear in that order, the mass-squared matrix takes the supermatrix form [8]

$$\hat{m}^2 = \begin{pmatrix} \hat{A} & 0 \\ 0 & \hat{B} \end{pmatrix} + \begin{pmatrix} 0 & \hat{G} \\ \hat{G}^\dagger & 0 \end{pmatrix}. \quad (2.11)$$

For $\lambda = 0$ there are two distinct sectors of opposite normality. The effect of the normality operator is to change $(0, \frac{1}{2})$ representations into $(\frac{1}{2}, 0)$ representations and vice versa. Since $\Pi$ commutes with $\hat{m}^2$, it follows that $\hat{A} = \hat{B}$ and $\hat{G} = \hat{G}^\dagger$. Eigenstates with $\eta = (\pm)$ are eigenvectors of $\hat{A} \pm \hat{G}$. In what follows we will concentrate solely on the helicity zero sector, unless otherwise stated.

We now consider the most general statements that we can extract from the A-W sum rule. There are two ways to proceed [3]. One can work directly with the physical states, as one would in saturating a dispersion integral with a given number of resonances, or one can make use of the representation theory to build up physical particle states as sums of irreducible representations of $SU(2) \times SU(2)$. Both methods will be used in this paper. The matrix element of $X_i$ between two arbitrary $I = \frac{1}{2}$ states, $\alpha$ and $\beta$, of opposite normality is

$$\langle\beta|X_i|\alpha\rangle = T_i\sqrt{\frac{4}{3}}\langle\beta||X||\alpha\rangle \equiv T_i g_{\beta\alpha\pi}, \quad (2.12)$$

---

3 Consider the $O(4)$ vector: $(2\bar{\psi}i\gamma_5 T_i\psi, \bar{\psi}\psi)$ [18].

where $T_i = \tau_i/2$ and the $\tau_i$ are the Pauli matrices, and $\langle\beta||X||\alpha\rangle$ is a reduced matrix element. Eq. (2.12) is simply the Wigner-Eckart theorem for a system of $I = \frac{1}{2}$ states. The generalized A-W sum rule, Eq. (2.5), can easily be shown to take the form [3]

$$\sum_\gamma g_{\beta\gamma\pi} g_{\gamma\alpha\pi} = \delta_{\beta\alpha}. \tag{2.13}$$

This form is ideally suited for a direct confrontation with experiment, as we will see below.

Next we establish the representation theory for general $I = \frac{1}{2}$ states. The action of $\Pi$ and $X_i$ on states labelled by their $SU(2)_L$ and $SU(2)_R$ content, respectively, is given by

$$\begin{aligned} \Pi|0\,\tfrac{1}{2}\rangle_l &= |\tfrac{1}{2}\,0\rangle_l \qquad (\Pi^2 = 1) \\ X_i|0\,\tfrac{1}{2}\rangle_l &= T_i|0\,\tfrac{1}{2}\rangle_l \\ X_i|\tfrac{1}{2}\,0\rangle_l &= -T_i|\tfrac{1}{2}\,0\rangle_l. \end{aligned} \tag{2.14}$$

The states of definite normality —labelled by their $\eta$ content— are

$$|\pm\rangle_l \equiv |0\,\tfrac{1}{2}\rangle_l \pm |\tfrac{1}{2}\,0\rangle_l. \tag{2.15}$$

Clearly $\Pi|\pm\rangle_l = \pm|\pm\rangle_l$ and $X_i|\pm\rangle_l = T_i|\mp\rangle_l$. We can now express the physical states, $\beta$ and $\alpha$, as sums of any number of fundamental states belonging to the allowed irreducible representations:

$$\begin{aligned} |\beta\rangle &= \sum_k b_k |-\rangle_k \\ |\alpha\rangle &= \sum_l a_l |+\rangle_l, \end{aligned} \tag{2.16}$$

where $b_k$ and $a_l$ are unknown coefficients. Sandwiching $X_i$ between these states yields

$$\langle\beta|X_i|\alpha\rangle = T_i \frac{\sum_k b_k a_k}{\sqrt{\sum_l |b_l|^2 \sum_m |a_m|^2}}. \tag{2.17}$$

The Schwarz inequality then implies $|g_{\beta\alpha\pi}| \leq 1$. So the pion interactions of the helicity zero states of $I = \frac{1}{2}$ mesons are bounded as a consequence of their participation in multiplets of unbroken $SU(2) \times SU(2)$. In order to go further —that is, fix the coefficients, $b_k$ and $a_l$— we must further constrain the asymptotic behaviour of the pion scattering amplitude.

All that remains unconstrained is the part of the crossing-even amplitude which carries $I_t = 0$. Therefore, Regge trajectories with vacuum quantum numbers necessarily play a crucial role in determining the chiral multiplet structure of hadrons in the broken phase. Regge pole theory suggests

$$M^{I_t=0}_{\beta j, \alpha i}(\omega, \lambda) \xrightarrow[\omega\to\infty]{} \omega^{\alpha_0(0)}, \tag{2.18}$$

where $\alpha_0$ is the leading trajectory with vacuum quantum numbers. We assume that there are no $I = 0$ Regge trajectories with $\alpha_0(0) \geq 0$ which contribute to transitions in which $\alpha$ and $\beta$ represent different physical states [3,7]. In other words, we assume that scattering becomes purely elastic at high energies. Certainly phenomena suggest that this a good approximation. For example, the cross sections for the processes $\pi + N \to a_1(1260) + N$ and $N + N \to N^*(1440) + N$ are less than 10% of those for $\pi + N \to \pi + N$ and $N + N \to N + N$, respectively [19,20]. Although, this can be understood on the basis of simple "diffraction" arguments, it is certainly not clear —from the point of view of QCD— why this is the case. Here we treat this constraint as experimental input. However, there also exists compelling theoretical evidence in favor of this assumption. In the pion sector, this assumption (together with the other algebraic sum rules) leads to ubiquitous relations such as $m_\rho^2 = 2g_{\rho\pi\pi}^2 f_\pi^2$ —known as one of the KSRF relations [21], and $m_{a_1}^2 = 2m_\rho^2$, which can be derived independently using spectral function sum rules and vector meson dominance [22]. In any case, this assumption leads to a superconvergence relation, which can be expressed in Lie-algebraic form as [3,7]

$$[m^2, [X_i^\lambda, [X_j^\lambda, m^2]]]_{\beta\alpha} = 0. \qquad (2.19)$$

This sum rule is simply the statement that $\hat{m}_0^2$ and $\hat{m}_4^2$ commute, or using Eq. (2.10), $\langle\alpha|\hat{m}_4^2|\beta\rangle = 0$ when $\alpha \neq \beta$. We can immediately extract the general consequences of this sum rule for the helicity zero states of the $I = \frac{1}{2}$ mesons. We specialize a general theorem proved in Ref. 8 to zero helicity.

As shown above, physical eigenstates of $\eta = (\pm)$ are eigenstates of $\hat{A} \pm \hat{G}$. Eq. (2.19) implies that $\hat{A}$ and $\hat{G}$ commute. Suppose that the vector $\vec{a}$ represents a physical state in the $\eta = (+)$ basis, as in Eq. (2.16). Since $\hat{A}$ and $\hat{G}$ commute, $\vec{a}$ is a simultaneous eigenvector of $\hat{A}$ and $\hat{G}$, say with eigenvalues $\mu^2$ and $\Delta$, respectively. Similarly, suppose that the vector $\vec{b}$ represents a physical state in the $\eta = (-)$ basis; $\vec{b}$ is also a simultaneous eigenvector of $\hat{A}$ and $\hat{G}$. There are then two possibilities: $\vec{a}$ and $\vec{b}$ have different eigenvalues in which case $\vec{a} \cdot \vec{b} = 0$, or $\vec{a}$ and $\vec{b}$ have the same eigenvalues in which case $|\vec{a} \cdot \vec{b}| = 1$. Therefore, if we assume that there are no degenerate states of the same normality, it follows that only pairs of states with masses $\mu^2 \pm \Delta$ communicate by single-pion emission and absorption. Since $\vec{a}$ and $\vec{b}$ are the physical states defined in Eq. (2.16), it follows immediately that $|g_{\beta\alpha\pi}| = 1$ if $\alpha$ and $\beta$ are paired, or $g_{\beta\alpha\pi} = 0$, otherwise. This result is completely general and applies to all $\lambda = 0$ states of the $I = \frac{1}{2}$ mesons[4]. These mesons come in two varieties: Kaons and heavy mesons ($D$- and $B$-mesons). The Kaons are best suited to phenomenological analysis, and will be investigated later. However, the interactions of heavy mesons with pions are also constrained by heavy quark symmetry.

---

4 In what follows, when we refer to the algebraic content of chiral symmetry, or the consequences of unbroken chiral symmetry, we will have this general consequence of the *three* chiral commutation relations in mind.

## 3. Heavy Quark Symmetry

In general, it is problematic to study pion-meson interactions in a systematic lagrangian formulation. For example, the $\rho$ meson can decay to two pions inside any Feynman diagram in which it appears. This decay generates a momentum transfer of order the $\rho$ mass, and is therefore not consistent with chiral power counting. So in studying the representations of unbroken chiral symmetry involving $\rho$, one loses touch with the systematic effective lagrangian language[5]. In order to specialize to the subset of all chiral tree graphs — which was essential in deriving the algebraic sum rules— one must fall back on the large-$\mathcal{N}_c$ approximation [5,7]. When heavy mesons are involved, the situation is somewhat ameliorated since heavy quark symmetry implies helpful degeneracies. In particular, we can justify the chiral tree approximation without invoking the large-$\mathcal{N}_c$ approximation if we are willing to assume that the mass splitting between any two members of a given heavy meson chiral multiplet is small compared to $\Lambda_\chi$. As long as this is the case, restriction to chiral tree graphs is automatic when one works to leading order in $\chi PT$. Of course in $\chi PT$ there are tree graphs with any number of derivatives acting at the vertices, but these —together with loop graphs— are higher order in the chiral expansion, and therefore constitute a small effect. This observation serves to illustrate that with heavy mesons it is not clear that the large-$\mathcal{N}_c$ limit is necessary in order to justify the chiral tree graph approximation.

The effective lagrangian technology which allows one to study the chiral invariant pion transitions of heavy mesons is called heavy hadron $\chi PT$ [24]. Heavy hadron $\chi PT$ provides an expansion in powers of momenta divided by $\Lambda_\chi \sim m_\rho$, and in powers of $\Lambda_{QCD}$ divided by the heavy hadron mass. Hence, one would expect good results for charm and bottom mesons. Mesons containing a heavy quark can be classified by the spin ($s_\ell$) and the parity ($\pi_\ell$) of the light quark [14]. Consequently, heavy mesons fall into degenerate doublets labelled by $s_{\pm}^\pi = (s_\ell \pm \frac{1}{2})^{\pi_\ell}$. The ground state mesons have $s_\ell = \frac{1}{2}$ and $\pi_\ell = (-)$ and are denoted $P$ $(0^-)$ and $P^*$ $(1^-)$[6]. This doublet can be arranged [24] into the "superfield"

$$H_a = \frac{(1+\slashed{v})}{\sqrt{2}}\{P_a^{*\mu}\gamma_\mu - P_a\gamma_5\}, \tag{3.1}$$

which transforms as $H_a \to H_a U^\dagger_{ba}$ under $SU(3) \times SU(3)$. The first excited states have $s_\ell = \frac{1}{2}$ and $\pi_\ell = (+)$, and are denoted $P_0^*$ $(0^+)$ and $P_1'$ $(1^+)$. These states are unobserved in both the $D$ and $B$ meson systems, and yet necessarily play an important role in what follows. In fact, our results will provide a plausible explanation for why these states are not observed. At the next level we have $s_\ell = \frac{3}{2}$ and $\pi_\ell = (+)$, corresponding to $P_1$ $(1^+)$ and $P_2^*$ $(2^+)$. The states belonging to this multiplet are identified as $D_1(2420)$ (neutral) and $D_2^*(2465)$ (neutral), and $B_1(5725)$ and $B_2^*(5737)$. These four excited states can also be assembled [25,26] into the "superfields"

$$S_a = \frac{(1+\slashed{v})}{\sqrt{2}}\{P_{1a}'^\mu\gamma_\mu\gamma_5 - P_{0\,a}^*\}, \tag{3.2}$$

and

$$T_a^\mu = \frac{(1+\slashed{v})}{\sqrt{2}}\{P_{2\,a}^{*\mu\nu}\gamma_\nu - \sqrt{\frac{3}{2}}P_{1a}^\nu\gamma_5[g_\nu^\mu - \frac{1}{3}\gamma_\nu(\gamma^\mu - v^\mu)]\}. \tag{3.3}$$

---

5   This is true only for the $\lambda = 0$ sector. Recently, it has been shown that one can study pion transitions within the vector nonet in systematic fashion, since mass splittings are of order the pion mass or less [23]. Here unbroken chiral symmetry gives an interesting prediction: if the $\lambda = 1$ states of the $\rho$ and the $\omega$ fill out a (2,2) representation of $SU(2) \times SU(2)$, then $|g_2| = 1$ (see also Ref. 4).

With respect to chiral symmetry, $S_a$ and $T_a^\mu$ transform like $H_a$. The three superfields transform as $H \to SH$ under the heavy quark spin symmetry group, $SU(2)_v$ $(S \in SU(2)_v)$. The kinetic terms of the three lowest lying doublets take the form

$$\begin{aligned}\mathcal{L}_{kin} &= -Tr[\bar{H}_a iv \cdot D_{ba} H_b] \\ &+ Tr[\bar{S}_a(iv \cdot D_{ba} - \delta m_S \delta_{ba})S_b] + Tr[\bar{T}_a^\mu(iv \cdot D_{ba} - \delta m_T \delta_{ba})T_{b\mu}],\end{aligned} \tag{3.4}$$

where mass splitting between the doublets has been taken into account. The residual masses, $\delta m_S = M_{P_0^*} - M_P = M_{P_1'} - M_P$ and $\delta m_T = M_{P_1} - M_P = M_{P_2^*} - M_P$, are defined in the heavy quark limit where the doublets are degenerate [25]. The covariant derivative is given by

$$D_{ab}^\mu \equiv \delta_{ab}\partial^\mu + V_{ab}^\mu = \delta_{ab}\partial^\mu + \frac{1}{2}(\xi^\dagger\partial^\mu\xi + \xi\partial^\mu\xi^\dagger)_{ab}, \tag{3.5}$$

and transforms as $(D_\mu H)_a \to (D_\mu H)_a U^\dagger_{ba}$ under $SU(3) \times SU(3)$. The Goldstone bosons are contained in the matrix field $\xi = \exp(i\mathcal{M}/f_\pi)$, where

$$\mathcal{M} = \begin{bmatrix} (1/\sqrt{2})\pi^0 + (1/\sqrt{6})\eta & \pi^+ & K^+ \\ \pi^- & -(1/\sqrt{2})\pi^0 + (1/\sqrt{6})\eta & K^0 \\ K^- & \bar{K}^0 & -\sqrt{\frac{2}{3}}\eta \end{bmatrix}. \quad (3.6)$$

---

6 The "$*$" superscript indicates positive normality ($\eta = (+)$), since, by convention, "$*$" is assigned to states in the spin-parity series $J^P = 0^+, 1^-, 2^+, \ldots$.

Out of $\xi$ one can also construct

$$A^\mu_{ab} = \frac{1}{2}i(\xi^\dagger \partial^\mu \xi - \xi \partial^\mu \xi^\dagger)_{ab}, \quad (3.7)$$

which transforms as $A^\mu{}_{ba} \to U_{ac} A^\mu{}_{cd} U_{db}$ under $SU(3) \times SU(3)$. The S-wave pion transitions within the heavy meson doublets are contained in the interaction lagrangian [24,25]

$$\mathcal{L}_{1\pi} = g\, Tr\left[\bar{H}_a H_b \slashed{A}_{ba} \gamma_5\right] + g'\, Tr\left[\bar{S}_a S_b \slashed{A}_{ba} \gamma_5\right] + g''\, Tr\left[\bar{T}_a T_b \slashed{A}_{ba} \gamma_5\right]. \quad (3.8)$$

The S-wave pion transitions among the heavy meson doublets arise from

$$\mathcal{L}_s = f'\, Tr\left[\bar{S}_a T^\mu_b A_{\mu ba} \gamma_5\right] + h\, Tr\left[\bar{H}_a S_b \slashed{A}_{ba} \gamma_5\right] + h.c. \quad (3.9)$$

Heavy quark symmetry forbids S-wave single-pion transitions between $T^\mu_a$ and the ground-state mesons [14]. In effect, this term is absent from $\mathcal{L}_s$. Higher partial waves arise from non-leading operators in the chiral expansion [25]. It is important to realize that these two lagrangians are fundamentally different. The mass splittings between the members of any heavy meson doublet are protected by heavy quark symmetry and therefore one might expect the interactions encoded in $\mathcal{L}_{1\pi}$ to respect chiral power counting. On the other hand, the mass splittings between heavy meson doublets are unbounded and so the interactions encoded in $\mathcal{L}_s$ are sensible only to the extent that these mass splittings are *observed* to be small. In the $D$ and $B$ meson systems the mass splitting between the $\frac{3}{2}^+$ doublet and the ground state doublet is around 400 MeV and so it is not unreasonable to expect that heavy hadron $\chi PT$ works for the interactions among the three lowest lying doublets.

Heavy quark symmetry also constrains the heavy meson mass matrix. Mass splitting within the ground state doublet arises from the chromomagnetic-moment interaction term [24]

$$\frac{\lambda_2}{m_Q} Tr\left[\bar{H}_a \sigma^{\mu\nu} H_a \sigma_{\mu\nu}\right], \quad (3.10)$$

which yields $M_{P^*} - M_P = -2\lambda_2/m_Q \equiv \Delta_2$. $\lambda_2$ is independent of the heavy quark mass —up to a logarithmic dependence which is calculable in perturbative QCD [27]. For charm and bottom mesons, the mass difference, $\Delta_2$, is of order the pion mass, or less. As noted, this small mass splitting is essential to the consistency of heavy hadron $\chi PT$. We learn more about the heavy meson mass matrix by working directly with the heavy quark degrees of freedom. The mass splitting operator in the heavy meson lagrangian arises from $1/m_Q$ corrections to the heavy quark lagrangian [28]

$$\mathcal{L}_{HQET} = \bar{Q} iv \cdot D Q + \frac{1}{2m_Q}\bar{Q}(iD)^2 Q - \frac{1}{2m_Q}\bar{Q}\sigma^{\mu\nu}(\frac{1}{2}gG_{\mu\nu})Q + \dots, \quad (3.11)$$

which is obtained from QCD in the $m_Q \to \infty$ limit. Here $Q$ represents a heavy quark field of definite velocity. The second and third term in Eq. (3.11) contribute to the heavy meson masses. The second term is the heavy quark kinetic energy and the third term arises from the chromomagnetic-moment of the heavy quark. The ground state heavy meson doublet contains a spin-zero and a spin-one meson. Hence, to order $1/m_Q$ the masses can be expressed as [29,30]

$$M_P = m_Q + \bar{\Lambda}(P) + \frac{1}{m_Q}\{\tilde{K}(P) + \tilde{G}(P)\} \quad (3.12)$$

$$M_{P^*} = m_Q + \bar{\Lambda}(P) + \frac{1}{m_Q}\{\tilde{K}(P) - \frac{1}{3}\tilde{G}(P)\}, \quad (3.13)$$

where $\bar{\Lambda}$ is a positive contribution —independent of the heavy quark mass— and

$$\tilde{K}(H) = \langle H|\frac{1}{2}\bar{Q}D^2Q|H\rangle \quad (3.14)$$

$$\tilde{G}(H) = \langle H|\frac{1}{2}\bar{Q}\sigma^{\mu\nu}(\frac{1}{2}gG_{\mu\nu})Q|H\rangle, \quad (3.15)$$

with $|H\rangle$ normalized to unity. These matrix elements are independent of the heavy quark mass. It then follows that $M_{P^*} - M_P = -\frac{4}{3}\tilde{G}/m_Q$, and one sees how the parameter $\lambda_2$ of heavy hadron chiral perturbation theory matches to a matrix element of a heavy quark operator. This general form for the masses in the $1/m_Q$ expansion implies useful constraints among linear combinations of masses. To order $1/m_Q$, the combination $3M_{P^*} + M_P$ is independent of $\tilde{G}$. This is a simple consequence of the fact that the leading mass splitting between $P$ and $P^*$ arises from the coupling of the spin of the light quark to the spin of the heavy quark. This constraint leads to relations among heavy hadron masses that agree well with experiment [29]. Applying identical arguments to the excited state doublets, one finds that the combinations $3M_{P'_1} + M_{P^*_0}$ and $3M_{P_1} + 5M_{P^*_2}$ are also independent of $\tilde{G}(H)$. In general, for a heavy meson doublet of states $P_{J_1}$ and $P^*_{J_2}$, the combination $(2J_1 + 1)M_{P_{J_1}} + (2J_2 + 1)M_{P^*_{J_2}}$ is independent of mass splitting between $P_{J_1}$ and $P^*_{J_2}$, to order $1/m_Q$.

One can also consider the constraints that exist among the squared masses. Directly from Eq. (3.12) and Eq. (3.13) we obtain

$$M_P^2 = (m_Q + \bar{\Lambda})^2 + 2\{\tilde{K} + \tilde{G}\} + O(\frac{1}{m_Q}) \tag{3.16}$$

$$M_{P^*}^2 = (m_Q + \bar{\Lambda})^2 + 2\{\tilde{K} - \frac{1}{3}\tilde{G}\} + O(\frac{1}{m_Q}), \tag{3.17}$$

and so $M_{P^*}^2 - M_P^2 = -\frac{8}{3}\tilde{G} + O(1/m_Q)$. Hence, in the heavy quark limit, the mass-squared splitting in the ground state heavy meson doublet is independent of the heavy quark mass (again up to logarithmic corrections). This is certainly borne out by experiment, which gives [31]:

$$\begin{aligned} M_{D^*}^2 - M_D^2 &= 0.56 \text{ GeV}^2 \\ M_{B^*}^2 - M_B^2 &= 0.55 \text{ GeV}^2 \\ M_{K^*}^2 - M_K^2 &= 0.53 \text{ GeV}^2. \end{aligned} \tag{3.18}$$

Of course, why this works for kaons as well remains a mystery. What constraints exist among linear combinations of squared masses? In the heavy quark limit, the combination $3M_{P^*}^2 + M_P^2$ is independent of $\tilde{G}$. In general, the combination $(2J_1+1)M_{P_{J_1}}^2 + (2J_2+1)M_{P_{J_2}^*}^2$ is independent of the mass-squared splitting between $P_{J_1}$ and $P_{J_2}^*$, in the heavy quark limit. In summary, constraints on the mass matrix, up to order $1/m_Q$, translate to identical constraints on the mass-squared matrix in the heavy quark limit. These constraints will prove useful in what follows. It is important to realize that these constraints are rigorous consequences of QCD in the heavy quark expansion.

## 4. Mended Chiral Symmetry in the Heavy Quark Limit

We are now in a position to consider the joint consequences of unbroken $SU(2) \times SU(2)$ and heavy quark symmetry. There is, of course, no loss of generality in considering the $SU(2) \times SU(2)$ representation content of the heavy mesons; heavy quark symmetry and $SU(3)$ automatically relate the isospin couplings to the strange couplings. We first relate the coupling constants that appear in the heavy hadron effective lagrangian to matrix elements of the axial-vector matrix, $X_i$:

$$\begin{aligned} \langle P|X_i^0|P^*\rangle &= gT_i \\ \langle P|X_i^0|P_0^*\rangle &= \langle P_1'|X_i^0|P^*\rangle = hT_i \\ \langle P_1'|X_i^0|P_0^*\rangle &= g'T_i \\ \langle P_1|X_i^0|P_0^*\rangle &= \langle P_1'|X_i^0|P_2^*\rangle = f'T_i \\ \langle P_1|X_i^0|P_2^*\rangle &= g''T_i. \end{aligned} \tag{4.1}$$

Here and below, the overall phases have been fixed by convention. In making this identification, we are clearly working to leading order in heavy hadron chiral perturbation theory. What about the transition matrix elements between non-zero helicity states? Since, to leading order in chiral perturbation theory, all single-pion transitions are in the S-wave, there is no distinction between the various transitions of definite helicity. For example, in general, the transition amplitude $P'_1 \to P^* + \pi$ has two independent coupling constants, and so the $\lambda = 0$ and $\lambda = \pm 1$ transitions are independent. However, when one restricts to S-waves, there is a single operator —proportional to $\epsilon(P'_1) \cdot \epsilon(P^*)$, where $\epsilon(H)_\mu$ is the polarization vector of $H$. The remaining transition amplitudes among the three lowest-lying heavy meson doublets are given by

$$\langle P^*|X_i^{\pm 1}|P^*\rangle = \pm g T_i$$
$$\langle P'_1|X_i^{\pm 1}|P^*\rangle = h T_i$$
$$\langle P'_1|X_i^{\pm 1}|P'_1\rangle = \pm g' T_i$$
$$\langle P'_1|X_i^{\pm 1}|P_1\rangle = \pm f' T_i \qquad (4.2)$$
$$\langle P'_1|X_i^{\pm 1}|P_2^*\rangle = f' T_i$$
$$\langle P_1|X_i^{\pm 1}|P_1\rangle = \pm g'' T_i$$
$$\langle P_2^*|X_i^{\pm 1}|P_2^*\rangle = \langle P_2^*|X_i^{\pm 2}|P_2^*\rangle = \pm g'' T_i.$$

Parity conservation relates the matrix elements with $\lambda$ to those with $-\lambda$ via the selection rule, Eq. (2.8). In the next section, we will see how the chiral representation content of the helicity zero states self-consistently determines the representation content of non-zero helicity states. This must be the case, as they are not independent, and thus provides an important consistency check. As shown above (see Eq. (2.17)), the heavy meson coupling constants are bounded as a consequence of the participation of heavy mesons in multiplets of unbroken chiral symmetry: $g$, $h$, $g'$, $f'$ and $g''$ must all take values between $-1$ and $1$.

Consider the representation of unbroken $SU(2) \times SU(2)$ involving the ground-state heavy meson doublet. As illustrated in Table 4.1, the general theorem deduced from the three chiral commutation relations allows two scenarios consistent with heavy quark symmetry: (i) $P$ is paired with $P_0^*$, and $P^*$ is paired with $P'_1$. This yields $|h| = 1$, $g = g' = f' = 0$. (ii) $P$ is paired with $P^*$. This yields $|g| = 1$ and $h = 0$. Here we assume that only adjacent heavy meson states participate in a given chiral multiplet. One might suppose that since $P_0^*$ and $P'_1$ are unobserved, $P$ could be paired with $P_2^*$. However, as noted above, this transition is forbidden by heavy quark symmetry. So in the absence of the $\frac{1}{2}^+$ doublet, case (ii) would have to be realized.

It is straightforward to show that case (ii) is inconsistent with QCD in the heavy quark limit. If $P$ and $P^*$ are paired, then we know from the general considerations discussed

**Table 4.1:** The "horizontal" heavy quark symmetry allows two general chiral multiplet structures. The brackets denote sectors that do not communitate by S-wave single-pion transitions.

$$\left\{\begin{array}{cc} P & P^* \\ \updownarrow & \updownarrow \\ P_0^* & P_1' \end{array}\right\} \qquad \{\ P \longleftrightarrow P^*\ \}$$

$$\left\{\begin{array}{cc} P_1 & P_2^* \\ & \\ & \vdots \end{array}\right\} \qquad \left\{\begin{array}{cc} P_0^* & P_1' \\ P_1 & P_2^* \\ & \vdots \end{array}\right\}$$

$$\text{(i)} \qquad\qquad\qquad \text{(ii)}$$

above that their squared-masses can be written as $\mu^2 \pm \Delta$, which leads to $M_{P^*}^2 - M_P^2 = 2\Delta$, where $\Delta$ ($\neq 0$) is a diagonal element of the matrix $\hat{m}_4^2$. This pairing is clearly inconsistent with the constraint that $3M_{P^*}^2 + M_P^2$ be independent of $\Delta$ in the heavy quark limit. Hence, case (ii) is inconsistent with QCD in the heavy quark limit[7]. On the other hand, as we will see below, case (i) easily accommodates this constraint. Since no two members of any heavy meson doublet have the same spin, no two members of any heavy meson doublet can be paired as in case (ii) and so our conclusion is universal; the trend exhibited by the two lowest lying doublets in Table 4.1(i) must be realized by all heavy meson *quartets* labelled by the light quark spin. Therefore, in the heavy quark limit, the algebraic content of chiral symmetry determines the S-wave[8] single-pion transition amplitudes of *all* heavy meson states. We exhibit the solution diagramatically in Figure 4.1. In a very general sense, this solution leads one to expect that for a given light quark spin, the members of the higher-lying heavy meson doublet should experience strong S-wave pion decays and therefore should have large widths (assuming, of course, kinematically open channels), and the members of the lower-lying heavy meson doublet should experience no S-wave pion decays and therefore should have small widths. As regards the lowest lying doublets, we conclude that $g = g' = g'' = f' = 0$ and $|h| = 1$. This is our main result.

---

7 Note that this argument implicitly assumes that there are no magical cancellations among the various terms which contribute to the heavy meson mass-squared matrix in the heavy quark limit. For example, if we fine-tune the mass matrix so that $-3\tilde{K} = \tilde{G}$, then presumably case (ii) would be allowed.

8 Here we are assuming that heavy hadron $\chi PT$ works not only within the lowest-lying quartet, but within any arbitrary quartet. That is, we assume that the mass splittings between heavy mesons with the same light quark spin are small compared to $\Lambda_\chi$.

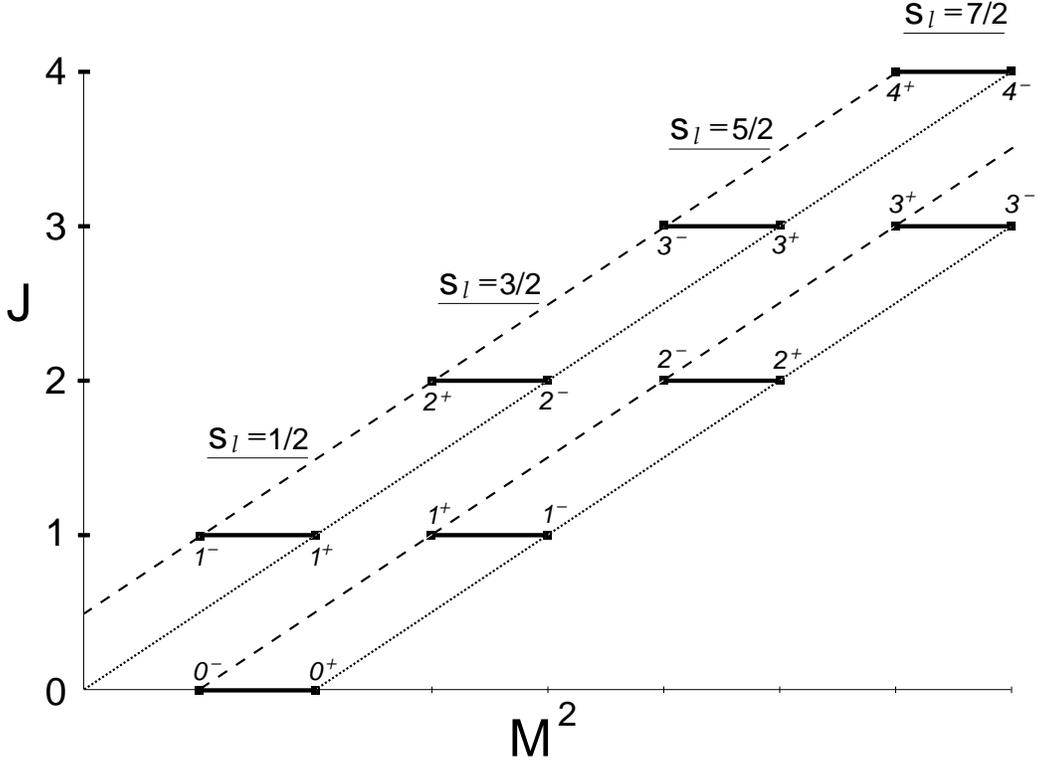

**Figure 4.1:** The eight lowest lying heavy meson doublets placed on Regge-like trajectories. Here we *assume* uniform mass-squared splittings between exactly degenerate doublets. The dashed (dotted) lines correspond to even (odd) normality trajectories. The solid lines denote S-wave single-pion transitions allowed by unbroken chiral symmetry.

As an aside, note that this conclusion also follows independently from a further constraint on the asymptotic behaviour of heavy meson scattering amplitudes. Again assume that $P$ and $P^*$ are paired. In the heavy quark limit, $P$ and $P^*$ are degenerate and so we must have $\Delta \ll M_P^2$. Suppose $\Delta = 0$ for simplicity. There is then no exchange of Regge trajectories with vacuum quantum numbers in the processes $\pi P \to \pi P$ and $\pi P^* \to \pi P^*$ in the heavy quark limit. There is no reason why this should be the case. In fact, this would suggest that the heavy quark limit is inconsistent with broken chiral symmetry. Put another way, chiral symmetry is spontaneously broken by the non-perturbative dynamics of the light quarks, independent of what the heavy quarks are doing, and so the heavy quark limit should not imply the vanishing of the matrix elements of $\hat{m}_4^2$ between physical states. That is, we expect that $\langle P_i | \hat{m}_4^2 | P_i \rangle \neq 0$ and $\langle P_i^* | \hat{m}_4^2 | P_i^* \rangle \neq 0$ when $M_{P_i^*}^2 = M_{P_i}^2$. The subscript labels the heavy quark doublet. This condition eliminates case (i) and yields the same conclusions as above. It is interesting that a QCD constraint on the heavy hadron mass-matrix is equivalent to a constraint on the exchange of Regge trajectories with vacuum quantum numbers in a heavy hadron scattering process.

Our work is not yet done. We must show that the heavy mesons can be accommodated in a representation of unbroken $SU(2) \times SU(2)$ which is consistent with the above constraints. There are, in principle, an infinite number of representations that will yield case (i). In general, $P$ is paired with $P_0^*$, with masses $\mu^2 \pm \Delta$, and $P^*$ is paired with $P_1'$, with masses $\nu^2 \pm \Delta'$. This corresponds to two independent chiral doublets. In the heavy quark limit a consistent solution to the QCD constraints exists if $\mu^2 = \nu^2$ and $\Delta' = \Delta/3$. We then have: $M_{P^*}^2 = \mu^2 + \frac{1}{3}\Delta$, $M_P^2 = \mu^2 - \Delta$, $M_{P_0^*}^2 = \mu^2 + \Delta$, and $M_{P_1'}^2 = \mu^2 - \frac{1}{3}\Delta$. These mass relations provide an interesting consistency check since evidently the algebraic content of chiral symmetry ensures that if $P$ and $P^*$ satisfy the QCD mass constraints, then so do $P_0^*$ and $P_1'$. Moreover, the relations between the mass-squared matrices of the chiral doublets imply that the doublets are not independent, but rather are embedded in a larger representation. We search for this representation in the next section.

## 5. Chiral quartets in the Heavy Meson Spectrum

Based on our assumptions, the ground- and first excited-state doublets do not communicate at all with higher level states via S-wave pion exchange. Therefore, we will consider a chiral quartet composed of $P$, $P^*$, $P_1'$, and $P_0^*$. The "true" representation could, in principle, be more complicated. However, as we will see, a quartet is the simplest representation which relates the squared-masses of these states in the manner required by unbroken chiral symmetry in the heavy quark limit. We will see how the various scenarios emerge in the representation theory. It is straightforward to show that the generalized A-W sum rule, Eq. (2.13), implies the three independent relations

$$\begin{aligned} g^2 + h^2 &= 1 \\ gh + hg' &= 0 \\ g'^2 + h^2 &= 1. \end{aligned} \qquad (5.1)$$

All S-wave transitions between the quartet and all higher excited states necessarily vanish, so $f' = 0$, *etc.* How do these sum rules arise in the representation theory? We can define four fundamental (normalized) states of definite normality:

$$|\phi_\pm\rangle \equiv \frac{1}{\sqrt{2}}(|0\ \tfrac{1}{2}\rangle_\phi \pm |\tfrac{1}{2}\ 0\rangle_\phi)$$
$$|\psi_\pm\rangle \equiv \frac{1}{\sqrt{2}}(|0\ \tfrac{1}{2}\rangle_\psi \pm |\tfrac{1}{2}\ 0\rangle_\psi). \tag{5.2}$$

The most general reducible chiral quartet filled out by the ground- and first excited-state heavy meson doublets is then

$$|P\rangle = \sin\theta|\phi_-\rangle - \cos\theta|\psi_-\rangle$$
$$|P_1'\rangle = \cos\theta|\phi_-\rangle + \sin\theta|\psi_-\rangle \tag{5.3}$$

$$|P^*\rangle = \sin\gamma|\phi_+\rangle + \cos\gamma|\psi_+\rangle$$
$$|P_0^*\rangle = \cos\gamma|\phi_+\rangle - \sin\gamma|\psi_+\rangle. \tag{5.4}$$

With this representation content, it follows from Eq. (4.1) that

$$g = -\cos(\theta + \gamma) = -g'$$
$$h = \sin(\theta + \gamma), \tag{5.5}$$

which are, of course, the entire content of the A-W sum rules, Eq. (5.1). Although here, we see that the values of the coupling constants that appear in the heavy meson lagrangian are fixed by a single unknown —the sum, $\theta + \gamma$, of the mixing angles between the fundamental states of definite normality.

What about the remaining non-zero helicity states? Within the quartet, there are two states with $\lambda = +1$, which can be put into the chiral doublet:

$$|P^*\rangle^{(+1)} = \sin\beta^+|0\ \tfrac{1}{2}\rangle_\delta - \cos\beta^+|\tfrac{1}{2}\ 0\rangle_\delta$$
$$|P_1'\rangle^{(+1)} = \cos\beta^+|0\ \tfrac{1}{2}\rangle_\delta + \sin\beta^+|\tfrac{1}{2}\ 0\rangle_\delta. \tag{5.6}$$

Taking matrix elements of $X_i$ between these states, and using Eq. (4.2) yields

$$g = -\cos(2\beta^+) = -g'$$
$$h = \sin(2\beta^+), \tag{5.7}$$

which again yield the entire content of Eq. (5.1), as they must. Therefore, $\beta^+ = (\theta + \gamma)/2$, and we see that the helicity zero states fix the representation content of non-zero helicity states, a consequence of the pure S-wave nature of the single-pion transitions to leading order in heavy hadron $\chi PT$. Notice also that the non-zero helicity states are then automatically consistent with the general theorem given in Ref. 8, which we specialized to zero helicity. Similarly, there are two states with $\lambda = -1$, which can be placed in the (independent) doublet:

$$|P^*\rangle^{(-1)} = \sin\beta^- |0\ \tfrac{1}{2}\rangle_\eta - \cos\beta^- |\tfrac{1}{2}\ 0\rangle_\eta$$
$$|P_1'\rangle^{(-1)} = \cos\beta^- |0\ \tfrac{1}{2}\rangle_\eta + \sin\beta^- |\tfrac{1}{2}\ 0\rangle_\eta.$$
(5.8)

Parity conservation relates the $\lambda = 1$ and $-1$ mixing angles: $\beta^- + \beta^+ = \pi/2$.

Next we consider the constraints on the mass-squared matrix. Working directly with physical states, the superconvergent sum rule, Eq. (2.19), can be expressed in the form [3]

$$\sum_\gamma g_{\beta\gamma\pi} g_{\gamma\alpha\pi} \left(M_\gamma^2 - M_\beta^2\right) = 0.$$
(5.9)

Our quartet is subject to the sum rules

$$gh(M_{P^*}^2 - M_{P_1'}^2) + hg'(M_{P_0^*}^2 - M_{P_1'}^2) = 0$$
$$gh(M_P^2 - M_{P_0^*}^2) + hg'(M_{P_1'}^2 - M_{P_0^*}^2) = 0,$$
(5.10)

which together with Eq. (5.1) imply

$$gh(M_P^2 - M_{P_1'}^2) = gh(M_{P^*}^2 - M_{P_0^*}^2) = 0.$$
(5.11)

In accord with the general theorem, there are three ways to satisfy these sum rules. If the same-normality states are degenerate, then the sum rules leave the coupling constants undetermined. This option is simply not realized in nature and so in what follows we will ignore it. We then have our familiar alternatives: (i) $g = 0$, which implies $|h| = 1$, or (ii) $h = 0$, which implies $g = -g' = 1$ or $-1$. We can also see how these alternatives arise in the representation theory.

In the quartet basis in which $|0\ \tfrac{1}{2}\rangle_\phi$, $|0\ \tfrac{1}{2}\rangle_\psi$, $|\tfrac{1}{2}\ 0\rangle_\phi$, and $|\tfrac{1}{2}\ 0\rangle_\psi$ appear in that order in the mass-squared matrix, $\hat{m}^2$, the mass-squared sub-matrices take the form

$$\hat{A} = \begin{pmatrix} m_a^2 & m_c^2 \\ m_c^2 & m_b^2 \end{pmatrix}, \qquad \hat{G} = \begin{pmatrix} \tilde{m}_d^2 & \tilde{m}_f^2 \\ \tilde{m}_f^2 & \tilde{m}_e^2 \end{pmatrix},$$
(5.12)

where the elements of $\hat{A}$ and $\hat{G}$ are undetermined. It is now straightforward to obtain the matrix elements in the physical basis, using Eq. (5.3), Eq. (5.4), and Eq. (5.12). The odd-normality ($\eta = (-)$) chiral-scalar mass-squared matrix elements are given by

$$\langle P|\hat{m}_0^2|P\rangle = \tfrac{1}{2}(m_a^2 + m_b^2) - \tfrac{1}{2}(m_a^2 - m_b^2)\cos 2\theta - m_c^2 \sin 2\theta$$
$$\langle P_1'|\hat{m}_0^2|P_1'\rangle = \tfrac{1}{2}(m_a^2 + m_b^2) + \tfrac{1}{2}(m_a^2 - m_b^2)\cos 2\theta + m_c^2 \sin 2\theta$$
$$\langle P|\hat{m}_0^2|P_1'\rangle = \tfrac{1}{2}(m_a^2 - m_b^2)\sin 2\theta - m_c^2 \cos 2\theta,$$
(5.13)

and the chiral-four-vector matrix elements are given by

$$\langle P|\hat{m}_4^2|P\rangle = -\frac{1}{2}(\tilde{m}_d^2 + \tilde{m}_e^2) + \frac{1}{2}(\tilde{m}_d^2 - \tilde{m}_e^2)\cos 2\theta + \tilde{m}_f^2 \sin 2\theta$$

$$\langle P_1'|\hat{m}_4^2|P_1'\rangle = -\frac{1}{2}(\tilde{m}_d^2 + \tilde{m}_e^2) - \frac{1}{2}(\tilde{m}_d^2 - \tilde{m}_e^2)\cos 2\theta - \tilde{m}_f^2 \sin 2\theta \qquad (5.14)$$

$$\langle P|\hat{m}_4^2|P_1'\rangle = -\frac{1}{2}(\tilde{m}_d^2 - \tilde{m}_e^2)\sin 2\theta + \tilde{m}_f^2 \cos 2\theta.$$

Similarly, the even-normality ($\eta = (+)$) chiral-scalar matrix elements are given by

$$\langle P^*|\hat{m}_0^2|P^*\rangle = \frac{1}{2}(m_a^2 + m_b^2) - \frac{1}{2}(m_a^2 - m_b^2)\cos 2\gamma + m_c^2 \sin 2\gamma$$

$$\langle P_0^*|\hat{m}_0^2|P_0^*\rangle = \frac{1}{2}(m_a^2 + m_b^2) + \frac{1}{2}(m_a^2 - m_b^2)\cos 2\gamma - m_c^2 \sin 2\gamma \qquad (5.15)$$

$$\langle P^*|\hat{m}_0^2|P_0^*\rangle = \frac{1}{2}(m_a^2 - m_b^2)\sin 2\gamma + m_c^2 \cos 2\gamma,$$

and the chiral-four-vector matrix elements are given by

$$\langle P^*|\hat{m}_4^2|P^*\rangle = \frac{1}{2}(\tilde{m}_d^2 + \tilde{m}_e^2) - \frac{1}{2}(\tilde{m}_d^2 - \tilde{m}_e^2)\cos 2\gamma + \tilde{m}_f^2 \sin 2\gamma$$

$$\langle P_0^*|\hat{m}_4^2|P_0^*\rangle = \frac{1}{2}(\tilde{m}_d^2 + \tilde{m}_e^2) + \frac{1}{2}(\tilde{m}_d^2 - \tilde{m}_e^2)\cos 2\gamma - \tilde{m}_f^2 \sin 2\gamma \qquad (5.16)$$

$$\langle P^*|\hat{m}_4^2|P_0^*\rangle = \frac{1}{2}(\tilde{m}_d^2 - \tilde{m}_e^2)\sin 2\gamma + \tilde{m}_f^2 \cos 2\gamma.$$

In order that $\hat{m}^2$ be diagonal in the physical basis, we must have

$$\frac{1}{2}(m_a^2 - m_b^2 - \tilde{m}_d^2 + \tilde{m}_e^2)\sin 2\theta - (m_c^2 - \tilde{m}_f^2)\cos 2\theta = 0$$
$$\frac{1}{2}(m_a^2 - m_b^2 + \tilde{m}_d^2 - \tilde{m}_e^2)\sin 2\gamma + (m_c^2 + \tilde{m}_f^2)\cos 2\gamma = 0. \qquad (5.17)$$

There are six unknown matrix elements and two unknown angles —subject to two constraints. It is clear that further assumptions must be made in order to extract any interesting information about the mass-squared matrix.

Assume that $\hat{m}_0^2$ and $\hat{m}_4^2$ commute. Since $\hat{A}$ and $\hat{G}$ transform as distinct irreducible representations of $SU(2) \times SU(2)$, they are linearly independent. The condition that $\hat{A}$ and $\hat{G}$ commute then implies $m_a^2 = m_b^2$ and $\tilde{m}_d^2 = \tilde{m}_e^2$ [9]. We then have

$$M_P^2 = m_a^2 - \tilde{m}_d^2 - (m_c^2 - \tilde{m}_f^2)\sin 2\theta$$
$$M_{P_1'}^2 = m_a^2 - \tilde{m}_d^2 + (m_c^2 - \tilde{m}_f^2)\sin 2\theta$$
$$M_{P^*}^2 = m_a^2 + \tilde{m}_d^2 + (m_c^2 + \tilde{m}_f^2)\sin 2\gamma \qquad (5.18)$$
$$M_{P_0^*}^2 = m_a^2 + \tilde{m}_d^2 - (m_c^2 + \tilde{m}_f^2)\sin 2\gamma$$

---

[9] We can also have the trivial solution $m_c^2 = \tilde{m}_f^2 = 0$. This yields two decoupled (independent) doublets; *e.g.*, $P$ paired with $P_0^*$, and $P^*$ paired with $P_1'$.

subject to the conditions $m_c^2 \cos 2\theta = m_c^2 \cos 2\gamma = \tilde{m}_f^2 \cos 2\theta = \tilde{m}_f^2 \cos 2\gamma = 0$. These conditions imply that $2\theta = \pi/2 + n\pi$ and $2\gamma = \pi/2 + m\pi$, where $m$ and $n$ are arbitrary integers. Defining $\ell \equiv m + n$, we then have

$$\theta + \gamma = \frac{\pi}{2}(\ell + 1), \tag{5.19}$$

and so we once again find our two separate cases of interest: (i) $\ell$ is even in which case $g = g' = 0$ and $|h| = 1$, and (ii) $\ell$ is odd in which case $h = 0$ and $g = -g' = 1$ or $-1$. These alternatives are, as they must be, the same as those obtained directly from the sum rule, Eq. (2.19), and from the general theorem. However, we have shown that case (ii) is ruled out in the heavy quark limit, since it implies mass-squared splittings within heavy meson doublets that do not respect QCD constraints. Therefore, we know that $\ell$ must be even. Suppose $\theta = \gamma = 3\pi/4$. We then have

$$\begin{aligned} M_P^2 &= (m_a^2 + m_c^2) - (\tilde{m}_d^2 + \tilde{m}_f^2) \\ M_{P_1'}^2 &= (m_a^2 - m_c^2) - (\tilde{m}_d^2 - \tilde{m}_f^2) \\ M_{P^*}^2 &= (m_a^2 - m_c^2) + (\tilde{m}_d^2 - \tilde{m}_f^2) \\ M_{P_0^*}^2 &= (m_a^2 + m_c^2) + (\tilde{m}_d^2 + \tilde{m}_f^2). \end{aligned} \tag{5.20}$$

It is straightforward to prove that $m_c^2 = 0$ is a further consequence of the QCD constraints on the mass-squared matrix in the heavy quark limit. From Eq. (5.20) it follows that

$$\begin{aligned} M_{P^*}^2 - M_P^2 &= -2m_c^2 + 2\tilde{m}_d^2 \equiv 4\delta_1 \\ M_{P_1'}^2 - M_{P_0^*}^2 &= -2m_c^2 - 2\tilde{m}_d^2 \equiv 4\delta_2 \\ 3M_{P^*}^2 + M_P^2 &= 4m_a^2 + 4(\delta_1 - \tilde{m}_f^2) \\ 3M_{P_1'}^2 + M_{P_0^*}^2 &= 4m_a^2 + 4(\delta_2 + \tilde{m}_f^2). \end{aligned} \tag{5.21}$$

In the heavy quark limit $m_a^2 = m_Q^2$ and so the general solution to the QCD constraints is $\delta_1 = -\delta_2 = \tilde{m}_f^2$, which gives $m_c^2 = 0$, as promised. This solution also implies $\tilde{m}_f^2 = \tilde{m}_d^2/2$. As one moves away from the heavy quark limit, this relation is necessarily badly broken as it implies that $P_1'$ lies below $P^*$, spectroscopically. This is not surprising, since $\tilde{m}_f^2$ is not protected by heavy quark symmetry; it governs the splitting between heavy meson doublets. Away from the heavy quark limit, $m_c^2$ can be non-zero. However, $m_c^2$ governs splitting within the heavy meson doublets and so is protected by heavy quark symmetry. Therefore, $m_c^2$ is small —of order $1/m_Q^2$ at the level of the mass matrix— in accord with observed mass-squared splittings (see Eq. (3.18)). We conclude that away from the heavy quark limit, the squared masses take the general form

$$\begin{aligned}
M_P^2 &= m_a^2 - (\tilde{m}_d^2 + \tilde{m}_f^2) \\
M_{P_1'}^2 &= m_a^2 - (\tilde{m}_d^2 - \tilde{m}_f^2) \\
M_{P^*}^2 &= m_a^2 + (\tilde{m}_d^2 - \tilde{m}_f^2) \\
M_{P_0^*}^2 &= m_a^2 + (\tilde{m}_d^2 + \tilde{m}_f^2),
\end{aligned} \quad (5.22)$$

which leads to the mass relation:

$$M_{P_0^*}^2 + M_P^2 = M_{P_1'}^2 + M_{P^*}^2. \quad (5.23)$$

This relation implies universal mass-squared splittings within the heavy meson doublets belonging to the lowest-lying quartet. Note also that in the $\hat{m}_4^2 \to 0$ limit there is complete degeneracy. Of course, our conclusions are general and apply to all heavy meson states. We conclude that the zero-helicity states of heavy mesons form reducible chiral quartets labelled by the light quark spin; generalizing Eq. (5.3) and Eq. (5.4) yields:

$$\begin{aligned}
&\frac{1}{\sqrt{2}}\{|\phi_- \; (s_\ell - \tfrac{1}{2}) \; -\rangle + |\psi_- \; (s_\ell - \tfrac{1}{2}) \; -\rangle\} \\
&-\frac{1}{\sqrt{2}}\{|\phi_- \; (s_\ell + \tfrac{1}{2}) \; +\rangle - |\psi_- \; (s_\ell + \tfrac{1}{2}) \; +\rangle\}
\end{aligned} \quad (5.24)$$

$$\begin{aligned}
&\frac{1}{\sqrt{2}}\{|\phi_+ \; (s_\ell + \tfrac{1}{2}) \; -\rangle - |\psi_+ \; (s_\ell + \tfrac{1}{2}) \; -\rangle\} \\
&-\frac{1}{\sqrt{2}}\{|\phi_+ \; (s_\ell - \tfrac{1}{2}) \; +\rangle + |\psi_+ \; (s_\ell - \tfrac{1}{2}) \; +\rangle\},
\end{aligned} \quad (5.25)$$

where the states are labelled by their $SU(2)_L \times SU(2)_R$, light quark spin, and parity content, respectively. The four lowest-lying quartets are exhibited in Figure 4.1.

Note the remarkable similarity between this representation and the representation involving the pion itself. There one also finds a quartet consisting of the helicity zero states of $\pi$, $\epsilon$ ($\sigma$ of old), $\rho$, and $a_1$ [32,3,7]. The representation content is as follows: $\epsilon$ and $1/\sqrt{2}(\pi + a_1)$ are in a $(2,2)$, $\rho$ is in a $(1,3) + (3,1)$, and $1/\sqrt{2}(\pi - a_1)$ is in a $(1,3) - (3,1)$, where the representations are labelled by their $SU(2)_L$ and $SU(2)_R$ content, respectively. This similarity in chiral multiplet structure offers a means of explaining why certain relations derived for heavy hadrons also work well for light hadrons. The relation between the mass-squared matrices of the heavy and light mesons will be considered in a separate work [33].

## 6. Heavy Meson Phenomenology

As noted above, the strong pion transitions of the heavy meson states are not well determined experimentally. Here we will see how our predictions compare with existing measurements. Much effort has centered on determining the $P^* \to P\pi$ ($P$ is a $D$ or a $B$ meson) transition amplitude. Besides being of interest in its own right, this matrix element appears in many hadronic form factors relevant to weak decays. For example, at small pion momentum, the $B^*$ pole dominates the semileptonic decay $B \to \pi e \bar{\nu}_e$ [34]. Here we will focus on the D meson system.

To leading order in chiral perturbation theory, the decay $D^* \to D\pi$ is determined by $g$ [24]:

$$\Gamma\left(D^{*+} \to D^0 \pi^+\right) = \frac{g^2}{12\pi f_\pi^2} |\vec{p}_\pi|^3. \qquad (6.1)$$

In our convention, $f_\pi = 93$ MeV. The decay width with a neutral pion in the final state is reduced by a factor of $1/2$. The anomalously small pion momentum ($|\vec{p}_\pi| = 40$ MeV) leads one to expect that even if $g = 1$ (as suggested by the constituent quark model [35]), the $D^*$ lifetime is long and so should be hard to measure. At present, there are lower and upper bounds on this decay. The experimental upper limit on the $D^*$ width [36] yields $g^2 < 0.5$. The radiative $D^*$ decays offer an indirect method of determining $g$, and lead to the lower bound $g^2 \gtrsim 0.1$ [35]. This lower bound is not necessarily in conflict with our prediction, as our results were obtained in the heavy quark and chiral limits. We will discuss each type of symmetry breaking in turn.

One might worry that heavy quark symmetry breaking could alter the chiral representation content of the heavy meson states, and thereby change the values of the coupling constants that were obtained in the heavy quark limit. This possibility is unlikely. Recall the discrete nature of the solution to the chiral commutation relations; the algebraic content of chiral symmetry implies that the single-pion transition amplitudes of $I = \frac{1}{2}$ states can take only three values (e.g. $-1$, $0$, or $1$ with suitable normalization), independent of heavy quark content or helicity [8]. Since the mixing angles which fix the representation content —and therefore the values of the coupling constants— are determined by the properties of the leading Regge trajectories with vacuum quantum numbers, the properties of these trajectories would have to be sensitive to the heavy quark mass, which we have already argued is not the case. In fact, we found that this argument is equivalent to QCD constraints on the heavy meson mass-squared matrix. Therefore, the solution —found in the heavy quark limit— should hold *order-by-order* in the $1/M$ expansion. This implies an infinite set of relations among symmetry breaking parameters in the heavy meson effective theory.

The leading heavy quark symmetry breaking corrections to the decay $P^* \to P\pi$ have recently been studied [37]. The relevant operators are

$$\frac{g_1}{M} Tr\left[\bar{H}_a H_b \slashed{A}_{ba} \gamma_5\right] + \frac{g_2}{M} Tr\left[\bar{H}_a \slashed{A}_{ba} \gamma_5 H_b\right], \tag{6.2}$$

where $g_1$ and $g_2$ are undetermined parameters. The effective coupling constants become [37]

$$\tilde{g}_P = g + \frac{(g_1 - g_2)}{M} \tag{6.3}$$

for the transition $P^* \to P\pi$, and

$$\tilde{g}_{P^*} = g + \frac{(g_1 + g_2)}{M} \tag{6.4}$$

for the transition $P^* \to P^*\pi$. Therefore, if the solution to the chiral commutation relations is indeed stable at each order in the $1/M$ expansion, it predicts $g = 0$ *and* $g_1 = g_2 = 0$. The latter prediction is, in principle, testable since these parameters contribute to any heavy meson process which receives corrections from pion loop graphs [37]. We will see below that there is important experimental evidence which suggests that the solution to the chiral commutation relations is stable to heavy quark symmetry breaking effects. In any case, as an exercise, we can easily check whether $g = 0$ is consistent with the experimental upper bound on $\tilde{g}$ in the $D$ meson system. Suppose $\tilde{g}^2 = 0.3$. We then have $|g_1 - g_2|/M_D = 0.55$. If, for example, $g_1 = -g_2$, we obtain $|g_1| \simeq 0.5$ GeV, which is consistent with the dimensional estimate: $|g_1|, |g_2| \sim \sqrt{\Delta_2 M_D} \simeq 0.5$ GeV.

There are two sorts of chiral symmetry breaking effects that we have to consider. In the decay width formula for $P^* \to P\pi$ we have used physical pions to compute kinematical factors, and yet the coupling constant $g$ was evaluated at zero pion mass; a non-zero pion mass interferes with the counting of powers of energy which was essential in deriving the Lie-algebraic sum rules [3]. Unfortunately, we know of no way of accounting for this small effect in a systematic fashion. The second type of breaking arises from chiral symmetry breaking operators in the effective lagrangian [38,39]. These effects lead to an effective coupling constant of the form

$$g_{eff} = g\left\{1 + O\left(\frac{m_\pi^2}{\Lambda_\chi^2} \log \frac{m_\pi^2}{\mu^2}\right) + \cdots\right\}, \tag{6.5}$$

where $\mu$ is an arbitrary scale. These corrections arise from non-analytic (in $m_q$) one loop graphs constructed from the leading order operators. This sort of correction is generically large [38]; *i.e.*, a 20% effect. However, since there is an overall factor of $g$, these corrections are *weighted* by $g$, and so vanish together with the axial-vector source. Evidently, the prediction $g = 0$ is not subject to explicit chiral symmetry breaking effects of this type.

Therefore, deviations of $g$ from 0 should be due entirely to chiral symmetry breaking effects of the kinematic type discussed above. In other words, the solution unambiguously predicts that the transition amplitude for the process $P^* \to P\pi$ should be very close to 0, and therefore the decay width for the process $D^* \to D\pi$ should be very close to the experimental lower bound implied by the radiative decays. This is the most currently relevant prediction made by the algebraic content of chiral symmetry in the heavy meson system. This transition matrix element has been calculated in many models with no particular value favored. A Table listing theoretical predictions is given in Ref. 40.

Although the $\frac{3}{2}^+$ states have been observed in the $D$ and $B$ meson systems, the $\frac{1}{2}^+$ states have not been observed. Of course these states are expected to exist. The constituent quark model suggests that S-wave decays should be strong [41], leading to the expectation that these states are very broad. The decay widths of the excited $\frac{1}{2}^+$ states to the ground state doublet are given by [25]

$$\Gamma(P_0^* \to P\pi^-) = \frac{|h|^2}{4\pi f_\pi^2}|\vec{p}_\pi(P_0^*, P)|^3$$
$$\Gamma(P_1' \to P^*\pi^-) = \frac{|h|^2}{4\pi f_\pi^2}|\vec{p}_\pi(P_1', P^*)|^3, \qquad (6.6)$$

where

$$|\vec{p}_\pi(\alpha, \beta)|^3 = \left(\frac{M_\beta}{M_\alpha}\right)(M_\alpha - M_\beta)^2\left[(M_\alpha - M_\beta)^2 - m_\pi^2\right]^{1/2}. \qquad (6.7)$$

If one takes $M_{D_1'} = M_{D_0^*} = 2.4$ GeV as suggested by the quark model [41] one finds

$$\Gamma(D_0^* \to D\pi^-) = |h|^2 \,[980 \text{ MeV}]$$
$$\Gamma(D_1' \to D^*\pi^-) = |h|^2 \,[400 \text{ MeV}], \qquad (6.8)$$

and so *a priori* it is not surprising that these states are unobserved. These results are quite sensitive to the choice of the heavy meson masses, and yet it is gratifying that the general solution, $|h| = 1$, implies that these decays should take the *maximum* value allowed by unbroken chiral symmetry (of course there is no reason to believe that non-analytic chiral symmetry breaking corrections to these decays are small). The fact that these states are unobserved supports the conjecture that the general solution is independent of heavy quark symmetry breaking effects; the alternative solution, $h = 0$, would unambiguously predict that these states are narrow.

**Table 7.1:** Contributions to the A-W sum rule for $\pi K$ scattering.

| $\gamma$ | $g^2_{\gamma K\pi}$ | $\gamma$ | $g^2_{\gamma K\pi}$ |
|---|---|---|---|
| $K^*(1410)$ | $<0.02$ | $K^*_3(1780)$ | 0.03 |
| $K^*_0(1430)$ | 0.08 | $K^*_0(1950)$ | $<0.02$ |
| $K^*_2(1430)$ | 0.17 | $K^*_4(2045)$ | $<0.02$ |
| $K^*(1680)$ | 0.06 | $K^*_5(2380)$ | $<0.02$ |

## 7. The Kaon system

We have seen that unbroken chiral symmetry and heavy quark symmetry determine the S-wave pion transition amplitudes of heavy meson states. What about the Kaons? If there is no heavy quark symmetry, we can make use of phenomenological input in order to predict coupling constants using the A-W sum rule. The best known example of this phenomenological approach is the A-W sum rule for the determination of the nucleon axial-vector coupling, $g_A$, as discussed above. We saw that an accurate result is obtained by including the many possible intermediate states [2]. From the point of view of chiral symmetry, it is not surprising that there are so many possible states in $\pi N$ scattering, since these can form reducible combinations of any number of $(0, \frac{1}{2})$, $(\frac{1}{2}, 0)$, $(0, \frac{3}{2})$, $(\frac{3}{2}, 0)$, $(1, \frac{1}{2})$, and $(\frac{1}{2}, 1)$ irreducible representations of $SU(2) \times SU(2)$. This large number of possible representations makes it difficult to predict the pion transitions of the unflavored baryons using the algbraic sum rules [8].

We can test the "phenomenological" A-W sum rule in the Kaon system in order to learn something about the representations of unbroken $SU(2) \times SU(2)$ filled out by the low-lying kaons. Here, in order to justify neglect of the continuum we must invoke the large-$\mathcal{N}_c$ approximation. Consider the representation involving $K$ and $K^*$. As with $g_A$, we define a coupling constant $g_{K^*K\pi}$, which is determined from the sum rule by saturating with all possible intermediate states in $\pi K$ scattering. Since the intermediate states are sums of $(0, \frac{1}{2})$ and $(\frac{1}{2}, 0)$ representations, we might expect fewer relevant intermediate states than in pion-nucleon scattering. This expectation is borne out experimentally. The A-W sum rule for $\pi K$ scattering can be expressed as:

$$g^2_{K^*K\pi} = 1 - \sum_\gamma g^2_{\gamma K\pi}, \tag{7.1}$$

where the coupling constants are related to the transition matrix elements in Eq. (2.12). The decay width for the process $\alpha \to \beta + \pi_i$ is given by

$$\Gamma\left(\alpha \to \beta + \pi_i\right) = \frac{|\vec{p}_\pi|^3 |\langle\beta|X_i|\alpha\rangle|^2}{2\pi f_\pi^2 \left(2J_\alpha + 1\right)}, \qquad (7.2)$$

where $\vec{p}_\pi$ is the pion momentum. Each term in the sum in Eq. (7.1) can be taken from experiment via the formula

$$g_{\gamma K\pi}^2 = 8\pi f_\pi^2 \left(2J_\gamma + 1\right) \Gamma_\gamma^{tot} B\left(\gamma \to K\pi\right)/3|\vec{p}_\pi^\gamma|^3, \qquad (7.3)$$

where we have used Eq. (2.12). In Table 7.1 we exhibit all observed intermediate states in $\pi K$ scattering. The experimental widths are central values quoted by the particle data group [42]. There is substantial experimental uncertainty associated with some of these decays.

Keeping all intermediate states with an appreciable contribution ($>$ 0.02), we find $|g_{K^*K\pi}| \simeq 0.66$, which does not agree very well with the experimental value, $|g_{K^*K\pi}^{exp}| = 0.44$. According to our general conclusions, we would expect that $|g_{K^*K\pi}| = 1$ or $0$. The experimental value favors neither of these alternatives. However, the couplings of $K$ to the set of intermediate states are generically small (see Table 7.1), and so the A-W sum rule suggests a representation content where $|g_{K^*K\pi}| = 1$. Presumably the discrepancy arises from chiral symmetry breaking effects, which one would expect to be large, and/or the limitations of the large-$\mathcal{N}_c$ approximation. One might then wonder why this chiral representation content differs from that of the ground state heavy meson doublet. It is clear that the Goldstone nature of the Kaon requires that $K$ and $K^*$ be paired. Algebraic realizations of $SU(2) \times SU(2)$ accurately predict the coupling of $\rho$ to two pions (KSRF relation) [3,7]. If we consider algebraic realizations of $SU(3) \times SU(3)$, the solution involving the pseudoscalar and vector octets necessarily incorporates the $SU(2) \times SU(2)$ realization as a special case, and so *a priori* $K$ and $K^*$ —which belong to the $\pi$ and $\rho$ $SU(3)$ multiplets, respectively— *must* communicate by single-pion emission and absorption. More specifically, $SU(3)$ symmetry implies $|\langle\pi||X||\rho\rangle| = |\langle K||X||K^*\rangle|$. Algebraic realizations of $SU(2) \times SU(2)$ give $|\langle\pi||X||\rho\rangle| = 1$, which immediately yields $|g_{K^*K\pi}| = 1$, an interesting consistency check. With $K$ and $K^*$ paired, the S-wave transition amplitude of any $0^+$ state to the $\pi K$ channel necessarily takes the smallest value that unbroken chiral symmetry allows, namely zero. In fact, $K_0^*(1430)$ contributes very little to the A-W sum rule [10], and is a scalar with well established properties. On the other hand, we have seen that in the heavy meson system the decay amplitude of the lowest lying scalar ($P_0^*$) takes the largest value allowed by unbroken chiral symmetry, and, in fact, is not observed. This is completely analogous

---

10 Note that although $K_0^*(1430)$ decays primarily to the $\pi K$ channel, and has a substantial width ($\sim 287\ MeV$), the pion momentum is large ($\sim 620\ MeV$). See Table 7.1.

**Table 7.2:** The chiral structure of the helicity zero states of the lowest lying mesons of a given character. The arrows indicate the "allowed" single-pion transitions.

$$\left\{\begin{array}{ccc} \pi & \longleftrightarrow & \rho \\ \updownarrow & & \updownarrow \\ \epsilon & \longleftrightarrow & a_1 \end{array}\right\} \qquad \left\{\begin{array}{cc} P & P^* \\ \updownarrow & \updownarrow \\ P_0^* & P_1' \end{array}\right\} \qquad \left\{\begin{array}{ccc} K & \longleftrightarrow & K^* \\ & & \end{array}\right\}$$

$$(a) \qquad\qquad\qquad (b) \qquad\qquad\qquad (c)$$

to the evanescent $\epsilon$ in the $\pi$ system, whose width is fixed using identical assumptions to a value so large that one would not expect to observe it [3,7]. Hence, the algebraic content of chiral symmetry offers a generic explanation for why low-lying scalars are observed in certain systems, and not in others. We illustrate these conclusions in Table 7.2.

## 8. Summary and Conclusion

In chiral perturbation theory, one parametrizes strong interaction physics at low-energies in a manner consistent with broken chiral symmetry. In general, for scattering processes, the leading order results (low-energy theorems) are completely fixed by chiral symmetry. Higher orders involve undetermined parameters that must be taken from experiment. This method is predictive because there are fewer undetermined parameters than observables. We have seen that chiral symmetry also has algebraic content which constrains some of these parameters. Extraction of the algebraic consequences of chiral symmetry requires an interplay between theory and experiment. In general, one must make use of experimental input in the form of asymptotic constraints on pion scattering amplitudes. These constraints lead to sum rules which can be expressed in Lie-algebraic form. In this paper we have made use of the following observationally inspired sum rules:

**(I)** The generalized Adler-Weisberger sum rule,

$$[X_i, X_j] = i\epsilon_{ijk} T_k \tag{8.1}$$

is the statement that hadrons fill out representations of $SU(2) \times SU(2)$ in the broken phase.

**(II)** The second sum rule, Eq. (2.10), is the statement that the mass-squared matrix is the sum of a term which transforms as a chiral scalar, and a term which transforms as the fourth component of a chiral four-vector:

$$\hat{m}^2 = \hat{m}_0^2 + \hat{m}_4^2. \tag{8.2}$$

For general pion scattering processes, this sum rule follows from the assumption that exotic Regge trajectories are absent. However, for chiral multiplets composed solely of $I = \frac{1}{2}$ mesons, this sum rule requires no additional assumption beyond Eq. (8.1).

**(III)** A third sum rule, Eq. (2.19), implies that

$$\left[\hat{m}_0^2, \hat{m}_4^2\right] = 0. \tag{8.3}$$

This is the fundamental assumption; although it is phenomenologically inspired, *it is the sole constraint used in this paper which cannot be directly traced to a symmetry of QCD*. Nevertheless, it clearly has deep group theoretical significance as it fixes the angles which mix the various irreducible representations of $SU(2) \times SU(2)$ that make up the heavy meson states.

These three sum rules severely constrain the single-pion transitions of $I = \frac{1}{2}$ mesons, and yet do not specify the representations of unbroken $SU(2) \times SU(2)$ filled out by these states. Recall that in the context of constituent quarks, the chiral commutations relations provided no means of distinguishing between the solutions $|g_A| = 1$ and $g_A = 0$ [8]. However, additional constraints that heavy quark symmetry puts on the heavy meson spectrum unambiguously predict the single-pion transitions of the heavy meson states. The single-pion transition matrix elements are related to the couplings that appear in the heavy hadron effective lagrangian. At leading order there are only S-wave single-pion transitions among the heavy meson states, and so a single coupling constant determines the transitions of all helicity states of a given heavy meson. The fundamental predictions are: (i) the S-wave single-pion transitions between heavy mesons belonging to a single heavy meson doublet take the *minimum* value allowed by unbroken $SU(2) \times SU(2)$ and so should be very weak (for example, $g = 0$). (ii) the S-wave transitions between heavy meson doublets of the same light quark spin take the *maximum* value allowed by unbroken $SU(2) \times SU(2)$, and so should be strong (for example, $|h| = 1$). (iii) Predictions (i) and (ii) are stable order-by-order in the $1/M$ expansion. This overall picture is consistent with current experimental data in the $D$ and $B$ meson systems.

There is much further work to be done exploring the algebraic consequences of chiral symmetry. We reiterate that the algebraic realizations provide an interesting symmetry-based method for relating properties of heavy and light hadrons. Given an arbitary heavy meson state, there is uncertainty associated with the coupling of the spin of the light quark to the heavy quark constituent. However, in the heavy quark limit, the dynamics of the heavy quarks decouple from the problem, leaving behind only the dynamics of the light quarks. The resulting constraints on the spectrum determine fundamental properties of the $I = \frac{1}{2}$ $SU(2) \times SU(2)$ representations. With this information one can use the $SU(2) \times SU(2)$ representation theory to construct $I = 0$ and $1$ meson states out of the *fundamental* $I = \frac{1}{2}$ states, and, in turn, relate properties of light and heavy mesons. In particular, this method offers a promising means of understanding why certain mass relations (e.g. see Eq. (3.18)) derived for heavy hadrons also work well for light hadrons [33]. As regards the results presented in this paper, it would be interesting to see if our general solution can be expressed as a direct consequence of an enlarged algebra in which the generators of angular momentum participate [3]. For example, in this way one can include D-wave transitions —which are sub-leading in heavy hadron $\chi PT$— as well as S-wave transitions, in a formalism which relates the various helicities. From a technical standpoint, it would be interesting to have some understanding of the role of explicit chiral symmetry breaking effects of kinematic type, which are relevant in the derivation of the algebraic sum rules [3]. An extension to $SU(3) \times SU(3)$ would allow one to study the representations of unbroken chiral symmetry filled out by the low-lying $SU(3)$ multiplets. This representation content —which one would also expect to be of quartet form [7]— would predict transition matrix elements between the pseudoscalar and vector octets, as well as potentially interesting mass-squared relations. It would also be interesting to consider algebraic photo-pion sum rules for heavy mesons [8,17], pion transitions of heavy Baryons, and implications of the algebraic content of chiral symmetry for the finite-temperature chiral phase transition.

## Acknowledgements

I thank C. Greiner, S.B. Liao, S. Matinyan, Ulf-G. Meißner, B. Müller, C.D. Roberts, M. Strickland, and especially R.P. Springer and U. van Kolck, for valuable conversations and criticism. I am grateful to M. Luke and S. Weinberg for clarifying several important points. This work was supported by the U.S. Department of Energy (Grant DE-FG05-90ER40592).